\documentstyle[prl,aps,epsfig,calc]{revtex}         

\begin{document}
\catcode`\ä = \active \catcode`\ö = \active \catcode`\ü = \active
\catcode`\Ä = \active \catcode`\Ö = \active \catcode`\Ü = \active
\catcode`\ß = \active \catcode`\é = \active \catcode`\è = \active
\catcode`\ë = \active \catcode`\ô = \active \catcode`\ê = \active
\catcode`\ø = \active \catcode`\ò = \active \catcode`\í = \active
\defä{\"a} \defö{\"o} \defü{\"u} \defÄ{\"A} \defÖ{\"O} \defÜ{\"U} \defß{\ss} \defé{\'{e}}
\defè{\`{e}} \defë{\"{e}} \defô{\^{o}} \defê{\^{e}} \defø{\o} \defò{\`{o}} \defí{\'{i}}
\draft               
\newcommand{\ifm}{interferometer }
\newcommand{\stz}{$|0\,\hbar k\rangle$ }
\newcommand{\stt}{$|2\,\hbar k \rangle$ }
\newcommand{\stmt}{$|-2\,\hbar k \rangle$ }
\newcommand{\stf}{$|4\,\hbar k \rangle$ }
\newcommand{\stmf}{$|-4\,\hbar k \rangle$ }
\newcommand{\om}{\omega_{\rm rec}}
\newcommand{\li}{$^6$Li}
\newcommand{\na}{$^{23}$Na}
\twocolumn[\hsize\textwidth\columnwidth\hsize\csname
@twocolumnfalse\endcsname 

\title{Fifty-fold improvement in the number of quantum degenerate fermionic atoms} \vspace{-5mm}
\author{Z. Hadzibabic, S. Gupta,  C.A. Stan, C.H. Schunck, M.W. Zwierlein,
K. Dieckmann, and W. Ketterle}
\address{Department of Physics, MIT-Harvard Center for Ultracold
Atoms, and Research Laboratory
of Electronics, \\
MIT, Cambridge, MA 02139}
\date{\today}
\maketitle

\begin{abstract}
We have produced a quantum degenerate {\li} Fermi gas with up to
$7 \times 10^7$ atoms, an improvement by a factor of fifty over
all previous experiments with degenerate Fermi gases. This was
achieved by sympathetic cooling with bosonic {\na} in the $F=2$,
upper hyperfine ground state. We have also achieved Bose-Einstein
condensation of $F=2$ sodium atoms by direct evaporation.
\end{abstract}
\vskip1pc]

\narrowtext

Over the last few years, there has been significant progress in
the production of quantum degenerate atomic Fermi gases
($^{40}$K\cite{dema99,roat02} and
{\li}\cite{trus01,schr01,gran02,hadz02}) and degenerate Bose-Fermi
mixtures ($^7$Li-$^6$Li\cite{trus01,schr01},
{\na}-{\li}\cite{hadz02}, and $^{87}$Rb-$^{40}$K\cite{roat02}).
These systems offer great promise for studies of new,
interaction-driven quantum phenomena.  The ultimate goal is the
attainment of novel regimes of BCS-like superfluidity in a gaseous
system\cite{houb99,holl01,ohas02,hofs02}. The current efforts to
induce and study strong interactions in a Fermi
gas\cite{loft02,diec02,ohar02,rega03,ohar02_2,gehm03,rega03_2,bour03,gupt03,rega03_3}
are complemented with the ongoing efforts to improve fermion
cooling methods, which would lead to lower temperatures and larger
samples.

The main reason why studies of degenerate Fermi gases are still
lagging behind the studies of atomic Bose-Einstein condensates
(BECs), is the complexity of cooling methods. The Pauli exclusion
principle prohibits elastic collisions between identical fermions
at ultra-low temperatures, and makes evaporative cooling of
spin-polarized fermionic samples impossible. For this reason,
cooling of fermions must rely on some form of mutual or
sympathetic cooling between two types of distinguishable
particles, either two spin states of the same
atom\cite{dema99,gran02}, or two different
atoms\cite{roat02,trus01,schr01,hadz02}. A key element in fermion
cooling is the design of better ``refrigerators" for sympathetic
cooling.

In this Letter, we report the first production of degenerate Fermi
samples comparable in size with the largest alkali
BECs\cite{abos01latt}. We successfully cooled up to $7 \times
10^7$ magnetically trapped {\li} atoms to below half the Fermi
temperature ($T_F$). This is an improvement in atom number by a
factor of 50 over the largest previously reported Fermi
sea\cite{rega03_3}. Further, in samples containing up to $3 \times
10^7$ atoms, we observed temperatures as low as $0.05\,T_F$, the
lowest ever achieved. At these temperatures, the fractional
occupation of the lowest energy state differs from unity by less
than $10^{-8}$.

As in our previous work\cite{hadz02}, {\li} atoms were
magnetically trapped in the $F=3/2$, upper hyperfine ground state,
and sympathetically cooled by bosonic {\na}. The crucial
improvement was our achievement of forced evaporation of sodium in
the $|F,m_F\rangle = |2,+2\rangle$, upper hyperfine ground state,
producing large and stable BECs with up to $10^7$ atoms. This
allowed us to create a magnetically trapped {\na}-{\li},
Bose-Fermi mixture which is stable against spin-exchange
collisions at all densities, and dramatically boosted our fermion
atom number.

\begin{figure}[htbf]
\begin{center}
\vskip0mm \epsfxsize=80mm {\epsfbox{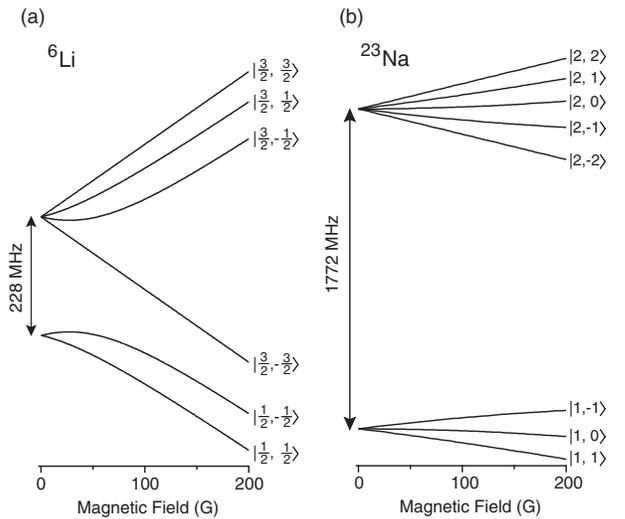}} \vskip1mm
\end{center}
\caption{Hyperfine structures of {\li} and {\na}. The states are
labelled in the low field, $|F, m_F\rangle$ basis. (a) Due to
finite trap depth of $\sim k_B \times 300\,\mu$K in the
$|1/2,-1/2\rangle$ state, lithium can be efficiently loaded into
the magnetic trap only in the upper, $F=3/2$ hyperfine state. (b)
Sodium is magnetically trappable in the $|1,-1\rangle$, and in the
$|F=2,m_F \geq 0\rangle$ states.  Previously, sodium has been
evaporatively cooled to BEC only in the $|1,-1\rangle$, lower
hyperfine state.} \label{fig1}
\end{figure}
\vskip-3mm

The criteria for designing sympathetic cooling experiments include
the heat capacity of the refrigerator, and the inter-species
collisional properties, both elastic and inelastic. Large and
stable {\na} condensates are an appealing choice for sympathetic
cooling of fermions. Further, a favorable mass ratio allows for
simultaneous Zeeman slowing of {\na} and {\li}\cite{hadz02}, and
for simultaneous magnetic trapping without large differences in
the gravitational sag. The inter-species collisional properties
are generally not predictable, and have to be tested
experimentally. In order to minimize all possible inelastic
processes, the natural choice is to magnetically trap both species
in their lower hyperfine ground states. However, at temperatures
reachable by laser cooling ($\geq 300\,\mu$K), {\li} can be
efficiently magnetically trapped only in the upper hyperfine
state, $F=3/2$ \cite{schr01,hadz02} (Fig.$\,$\ref{fig1}(a)). On
the other hand, until now sodium has been successfully evaporated
only in the lower, $F=1$ hyperfine state.  This was a limiting
factor for sympathetic cooling of {\li}, since the mixture of
sodium in the lower, and lithium in the upper hyperfine state is
not stable against spin-exchange collisions. The inelastic loss
rate increases as the temperature is lowered and the density
grows. In our previous work\cite{hadz02}, we partially overcame
this problem by transferring lithium atoms into the lower
hyperfine state after an initial sympathetic cooling stage to
$\sim 50\,\mu$K. By achieving forced evaporative cooling and
Bose-Einstein condensation of sodium in the $F=2$ state, we have
now realized a more robust sympathetic cooling strategy, and
dramatically improved the size and temperature of a degenerate
Fermi system.

We loaded $\sim 3 \times 10^9$ sodium and up to $10^8$ lithium
atoms in their upper hyperfine states from a two-species
magneto-optical trap (MOT) into the magnetic trap. The adverse
effect of light assisted collisions in a two-species
MOT\cite{hadz02,wipp02} was minimized by slightly displacing the
two MOTs with respect to each other. During the typical $30\,$s of
evaporative/sympathetic cooling, we observed no significant
inelastic loss of lithium atoms (by three-body collisions or
dipolar relaxation), the final number of degenerate atoms being at
least half of the number initially loaded into the trap. On the
other hand, we observed a favorable rate of elastic collisions
between the two species, with the inter-species thermalization
time being shorter than $1\,$s. Therefore, sodium atoms in the
upper hyperfine state have ideal properties as a refrigerant for
{\li}.

\begin{figure}[htbf]
\begin{center}
\vskip0mm \epsfxsize=80mm {\epsfbox{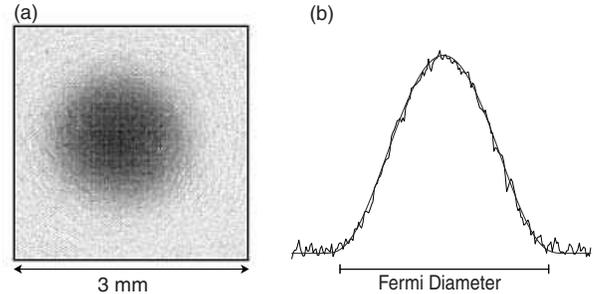}} \vskip1mm
\end{center}
\caption{Large and ultra-degenerate Fermi sea. (a) Absorption
image of $3\times 10^7$ {\li} atoms released from the trap and
imaged after $12\,$ms of free expansion.  (b) Axial (vertical)
line density profile of the cloud in (a). A semiclassical fit
(thin line) yields a temperature $T=93\,$nK$=0.05\,T_F$. At this
temperature, the high energy wings of the cloud do not extend
visibly beyond the Fermi energy, indicated in the figure by the
momentum-space Fermi diameter.} \label{fig2}
\end{figure}

Since our primary interest was cooling of fermions, we evaporated
all sodium atoms in order to get lithium to the lowest possible
temperatures. Even in our largest {\li} samples, of $\sim 7 \times
10^7$ atoms, we achieved temperatures below $0.5\,T_F$.
Temperatures in the range $0.05 - 0.2\,T_F$ could be achieved by
reducing the {\li} atom numbers only slightly, to $\sim 3 \times
10^7$. Such big clouds had a high enough optical density for crisp
absorption imaging even after ballistic expansion to a size larger
than one millimeter (Fig.$\,$\ref{fig2}(a)).

Temperatures were extracted from absorption images of expanding
clouds released from the trap, using a semiclassical
(Thomas-Fermi) fit to the Fermi-Dirac momentum distribution
\cite{hadz02,butt97} (Fig.$\,$\ref{fig2}(b)). The quoted
temperature range reflects both the shot-to-shot and day-to-day
reproducibility, and the fact that the Fermi distribution is very
insensitive to the temperature in this ultra-degenerate limit.

In these experiments, the {\li} atom number was adjusted during
the loading phase. Somewhat lower temperatures could possibly be
achieved if the maximum lithium atom number was loaded into the
magnetic trap, and then the hottest part of the cloud was
selectively removed by direct evaporation once the sodium atom
number dropped to the point where the heat capacities of the two
species become comparable. However, at this point it appears
unlikely that temperatures below $0.05\,T_F$ could be conclusively
extracted in order to differentiate the two strategies.

\begin{figure}[htbf]
\begin{center}
\vskip0mm \epsfxsize=80mm {\epsfbox{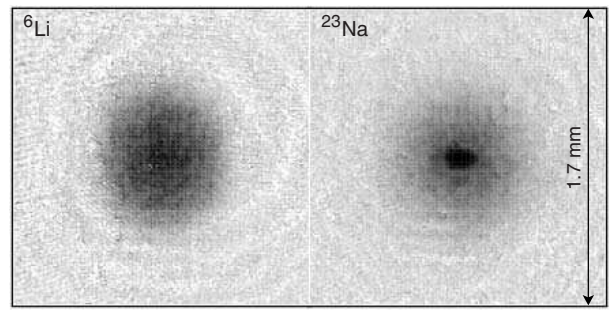}} \vskip1mm
\end{center}
\caption{Two-species mixture of degenerate Bose and Fermi gases.
After release from the magnetic trap, both {\li} and {\na} clouds
were imaged onto the same CCD camera using separate light pulses.
The times of free expansion of the two gases could be varied
independently. This dual-imaging technique allowed for optimizing
the cooling strategy for either single- or two-species
experiments. For the displayed image, the expansion times were
$\tau_{Li} = 8\,$ms and $\tau_{Na} = 25\,$ms, and the atom numbers
were $N_{Li} \sim 10^7$ and $N_{Na} \sim 6 \times 10^6$. Sodium
was cooled below the condensation temperature, corresponding to
$\sim 0.2\, T_F$ for the lithium cloud.} \label{fig3}
\end{figure}

We also produced two-species degenerate Bose-Fermi mixtures with
several million atoms in each species (Fig.$\,$\ref{fig3}). The
mixture was stable, with a lifetime of several seconds, limited
only by the three-body decay of the sodium cloud.

In the rest of the paper, we summarize the numerous steps which
were introduced to prepare sodium in the $F=2$ state as a
refrigerant.

In contrast to $^{87}$Rb, condensation of sodium by evaporative
cooling was previously achieved only in the lower, $|1,-1\rangle$
hyperfine state. $F=2$ sodium condensates could thus be studied
only by transferring optically trapped $F=1$ BECs into this
state\cite{gorl03,lean02_2}. Condensation in the upper hyperfine
state of sodium is more difficult than in the lower state for two
reasons: (1) The requirement for efficient optical pumping in
dense laser-cooled samples, and (2) an order of magnitude higher
three-body loss rate coefficient\cite{gorl03}.

The basic setup of our experiment is described in\cite{hadz02}. In
10$\,$s, we  collected typically $\sim 10^{10}$ {\na} atoms, and
$\sim 10^8$ {\li} atoms in a magneto-optical trap (MOT). Typical
MOT temperatures were 0.7-1$\,$mK. Sodium was collected in a
dark-SPOT variant of the MOT\cite{kett93}, and therefore most of
the atoms were in the $F=1$ hyperfine state. Lithium was collected
in a ``bright" MOT, with about 2/3 of the atoms in the $F=3/2$
state.

Before the transfer into the magnetic trap, the atoms were
optically pumped into the stretched hyperfine ground states,
$|2,+2\rangle$ for {\na}, and $|3/2,+3/2\rangle$ for {\li}. A
magnetic guide field of $3\,$G was applied, and the atoms were
optically pumped for $2\,$ms, using $\sigma^+$ polarized light. To
achieve both $F$ (hyperfine) and $m_F$ (Zeeman) pumping, two light
beams where used for each species, resonant with the $|F=I \pm
1/2\rangle \rightarrow |F'= I \pm 1/2\rangle$ transitions. Here,
$I$ is the nuclear spin ($I=3/2$ for {\na}, and $I=1$ for {\li}),
and $F'$ is the total spin in the excited electronic state. In
this way, almost all the lithium atoms could be pumped into the
$|3/2,+3/2\rangle$ state. On the other hand, the density of sodium
atoms in the dark-SPOT is $\geq 10^{11}\,$cm$^{-3}$, and Zeeman
pumping is notoriously difficult at such high densities. In our
experiments, the fraction of atoms pumped into the $|2,+2\rangle$
state was limited to about $30\,\%$, with most of the remaining
atoms distributed among the other $m_F$ sub-levels of the $F=2$
manifold.

After the optical pumping stage, the atoms were loaded into a
Ioffe-Pritchard magnetic trap with a radial gradient of
$164\,$G/cm, and axial curvature of $185\,$G/cm$^2$. Sodium atoms
in all three $|F=2, m_F \geq 0\rangle$ states are (at least
weakly) magnetically trappable (Fig.$\,$\ref{fig1}(b)). However,
only pure $|2,+2\rangle$ samples are stable against inelastic
spin-exchange collisions. A crucial step in preparing the samples
for efficient forced evaporation was to actively remove $|F=2,
m_F=0,+1\rangle$ atoms from the trap, before they engaged in
inelastic collisions with the $|2,+2\rangle$ atoms. The atoms were
loaded into a weak magnetic trap, with a high bias field of
$80\,$G. This field splits the $F=2$ Zeeman sub-levels by $\sim
k_B \times 2.8\,$mK. Since this splitting was larger than the
temperature of the cloud, the different states could be resolved
in microwave or rf spectroscopy, and the $|F=2, m_F=0,+1\rangle$
atoms could be selectively transferred to the untrapped $|F=1,
m_F=0,+1\rangle$ lower hyperfine states. This transfer was done
with a microwave sweep near the {\na} hyperfine splitting of
$1.77\,$GHz. In this way, all the $|2,+2\rangle$ atoms initially
loaded into trap could be preserved. We were even able to
``recycle" some of the untrapped atoms by optically pumping them
out of the $F=1$ ground states, thus giving them a ``second
chance" to fall into the $|2,+2\rangle$ state. The final setup
consisted of two microwave sweeps, the first of $0.8\,$s duration
with the optical pumping light on, and the second of $2.4\,$s
duration without the light. In this way, the overall transfer
efficiency from the MOT to the magnetic trap was improved to about
$35\,\%$, comparable to our standard $F=1$ BEC experiments
\cite{mewe96}.

After this purification of the $|2,+2\rangle$ sample, the magnetic
trap was tightened by reducing the bias field to $3.8\,$G in
2.4$\,$s. Resulting trapping frequencies were $204\,$Hz
($400\,$Hz) radially, and $34\,$Hz ($67\,$Hz) axially for the
sodium (lithium) stretched state. This provided good conditions
for forced runaway evaporation of sodium. Evaporation was done on
the $|2,+2\rangle \rightarrow |1,+1\rangle$ microwave transition
near $1.77\,$GHz. In contrast to radio-frequency evaporation, this
insured that {\li} was far off resonance.  Further, microwave
evaporation avoided any undesirable aspects of ``incomplete
evaporation" into the $|F=2, m_F=0,+1\rangle$ states, which could
lead to inelastic losses \cite{desr99}.

After $15\,$s of evaporation, the sodium atoms reached a
temperature of $T \sim 10\,\mu$K. At this point, to avoid
three-body losses in the $|2,+2\rangle$ state\cite{gorl03}, the
trap was weakened to frequencies of $49\,$Hz ($96\,$Hz) radially,
and $18\,$Hz ($35\,$Hz) axially for sodium (lithium). The final
evaporation to BEC took another $15\,$s. In this way, in the
absence of lithium atoms, we could produce almost pure
$|2,+2\rangle$ BECs containing up to 10 million atoms. The
lifetime of the BEC in the weak trap was longer than $3\,$s. In
contrast to our previous work\cite{gorl03,lean02_2}, studies of
$F=2$ condensates are now possible without the added complexity of
an optical trap.

In conclusion, by creating a superior refrigerant for sympathetic
cooling of {\li}, we have produced the coldest and by far the
largest quantum degenerate Fermi gas so far. With the number of
atoms comparable with the largest alkali BECs, and the
temperatures reaching the practical detection limit, we have fully
exploited the potential of laser and evaporative cooling to
engineer samples of ultracold fermions. In analogy with
Bose-Einstein condensates, we expect these large samples to insure
sufficient signal-to-noise ratio for all the standard techniques
of BEC research, such as velocimetry using long expansion times,
rf spectroscopy with Stern-Gerlach separation during ballistic
expansion, direct non-destructive imaging of the trapped clouds,
and Bragg spectroscopy.  The next challenge is to maintain a
similar combination of number and temperature for an interacting
two-component Fermi gas\cite{gupt03}.

We thank A. E. Leanhardt for critical reading of the manuscript.
This work was supported by the NSF, ONR, ARO, and NASA.

\end{document}